\documentclass[conference]{IEEEtran}
\IEEEoverridecommandlockouts
\usepackage{cite}
\usepackage{amsmath,amssymb,amsfonts}
\usepackage{algorithmic}
\usepackage{graphicx}
\usepackage{textcomp}
\usepackage{xcolor}
\usepackage{siunitx}
\usepackage{multirow}
\usepackage{tikz, tikzscale}
\usepackage{booktabs}
\usepackage{balance}
\usetikzlibrary{matrix,chains,positioning,decorations.pathreplacing,arrows,fadings}
\usepackage{pgfplotstable}
\usepackage{pgfplots}
\usepackage{comment}
\usepackage{hyperref}
\usetikzlibrary{pgfplots.statistics, pgfplots.colorbrewer}
\pgfplotsset{compat=1.9}
\makeatletter
\pgfplotsset{
    unit code/.code 2 args=
    \begingroup
    \protected@edef\x{\endgroup\si{#2}}\x
}
\DeclareRobustCommand\sampleline[1]{%
  \tikz\draw[#1] (0,0) (0,\the\dimexpr\fontdimen22\textfont2\relax)
  -- (2em,\the\dimexpr\fontdimen22\textfont2\relax);%
}
\def\BibTeX{{\rm B\kern-.05em{\sc i\kern-.025em b}\kern-.08em
    T\kern-.1667em\lower.7ex\hbox{E}\kern-.125emX}}
\begin{document}

\title{Interpreting End-to-End Deep Learning Models for Speech Source Localization Using Layer-wise Relevance Propagation\\

}

\author{\IEEEauthorblockN{ Luca Comanducci, Fabio Antonacci, Augusto Sarti}
\IEEEauthorblockA{\textit{Dipartimento di Elettronica, Informazione e Bioingegneria (DEIB), Politecnico di Milano} \\
Piazza Leonardo Da Vinci 32, 20133 Milan, Italy \\
name.surname@polimi.it}}

\maketitle

\begin{abstract}
Deep learning models are widely applied in the signal processing community, yet their inner working procedure is often treated as a black box. 
In this paper, we investigate the use of eXplainable Artificial Intelligence (XAI) techniques to analyze learning-based end-to-end speech source localization models. We consider the Layer-wise Relevance Propagation (LRP) technique, which aims to determine which parts of the input are more important for the output prediction. Using LRP we analyze two state-of-the-art models, of differing architectural complexity that map audio signals acquired by the microphones to the cartesian coordinates of the source. Specifically, we inspect the relevance associated with the input features of the two models and discover that both networks seem to denoise and de-reverberate the microphone signals in order to compute more accurate statistical correlations between them and consequently localize the sources. To further demonstrate this fact, we estimate the Time-Difference of Arrivals (TDoAs) via the Generalized Cross Correlation with Phase Transform (GCC-PHAT)  using both microphone signals and relevance signals extracted from the two networks and show that through the latter we obtain more accurate time-delay estimation results.
\end{abstract}


\section{Introduction}
The use of deep learning techniques has become ubiquitous in most fields of science, due to their ability to obtain more accurate results than the majority of the previous state-of-the-art methodologies. 
Following this trend, there has been an increasing adoption of deep learning techniques also in the acoustic signal processing field~\cite{cobos2022overview,bianco2019machine}, to tackle problems such as speech source localization~\cite{vera18,sundar2020raw,comanducci2020source}, DoA and TDoA estimation~\cite{chakrabarty2019multi,comanducci2020time}, speech separation~\cite{wang2018supervised} or 3D audio~\cite{morgado2018self}. 

The main drawback of deep learning techniques is their behavior as black-boxes, which gives the user very little control or understanding of the physical meaning, if any, of the nonlinear mappings learned by the network. This has motivated the surge of the eXplainable Artificial Intelligence (XAI) field, whose aim is to develop methods for interpreting, explaining, or visualizing features learned by deep learning models~\cite{samek2019towards}. The application of XAI techniques has become common practice in the wider machine learning community, but it is still lacking in most acoustic signal processing fields.
XAI has been widely adopted in image processing and computer vision
~\cite{zeiler2014visualizing, Simonyan2014, Mahendran2015Understanding, dosovitskiy2016inverting},  where the task is easier due to the presence of visual cues. A similar reasoning can be applied in the audio domain when considering frequency-based representations as input to the network, such as spectrograms~\cite{becker2024audiomnist, pellegrini2016inferring,comanducci2021reconstructing} or Head-Related Transfer Functions (HRTF)~\cite{thuillier2018spatial}. The adoption of visualization techniques in the case of raw audio, instead, is not straightforward. An early attempt was made in~\cite{muckenhirn2019understanding}, where a gradient-based approach was applied to study networks created for tasks such as speaker recognition and phone classification.

We consider the Layer-wise Relevance Propagation (LRP) technique~\cite{bach2015pixel}, an explainability technique that propagates the prediction of the network backward to the input, allowing to visualize which elements of the input data have been more important in determining the output. 
In~\cite{perotin2019crnn} LRP was used in order to extensively analyze the input Ambisonics acoustic intensity features to the source localization network for several noise/reverberation conditions, but it has never been applied to end-to-end localization networks, i.e. mapping raw audio signals to cartesian coordinates.

In this paper, we analyze two deep-learning architectures for end-to-end speech source localization. The one proposed in~\cite{vera18}, in the following \textit{LocCNN}, whose core structure consists of a simple sequence of 1D-convolutional layers, and the much more complex~\cite{sundar2020raw}, based on a \textit{SampleCNN}~\cite{lee2018samplecnn} architecture, which contains residual Squeeze-and-Excitation blocks~\cite{hu2018squeeze}. We first inspect the raw audio features and the corresponding relevance signals.
We then convert them into frequency-based representations, discovering that the networks act as a ``gate'' function, denoting as more relevant, samples corresponding to the onset of the speech signals received by the microphones. These observations suggest that the networks discard information related to the intelligible content of the speech, to focus only on the temporal information, as also suggested by listening to the relevance signals. To find out if this is true we inspect the GCC-PHAT~\cite{knapp1976generalized} and discover that the one computed using the relevance signals contains fewer spurious peaks than the one obtained via the microphone signals. We perform an experiment by estimating the TDoA over the whole test dataset using microphone and relevance signals. Results suggest that both networks learn to denoise and de-reverberate the signals in order to better correlate them and consequently estimate the source position. The rest of the paper is organized as follows. In Sec. \ref{sec:lrp} we present LRP and how it was applied to the LocCNN and SampleCNN models, described in Sec.~\ref{sec:setup} along with the simulated scenarios. In Sec.~\ref{sec:experiments} we present the experiments and results. Finally, Sec.~\ref{sec:conclusion} concludes the paper.
\section{Layer-wise Relevance Propagation}
\label{sec:lrp}
The LRP technique works by redistributing relevance values associated with neurons in upper layers backward to the input. This procedure is subject to a conservation property, meaning that the relevance received by a given layer must be redistributed to the lower layer in an equal amount~\cite{samek2019explainable}. The propagation of the relevance score $R_k$ at a given layer to a generic neuron $j$ from an immediately lower layer can be obtained by applying the rule
\begin{equation}
    R_j = \sum_k \frac{z_{jk}}{\sum_j z_{jk}}R_k,
\end{equation}
where $R_j$ corresponds to the relevance score at neuron $j$ due to neuron $k$. The quantity $z_{jk}$ models how much neuron $j$ has contributed to making neuron $k$ relevant, while the denominator enforces the conservation property. The way in which $z_{jk}$ is computed is denoted as \textit{rule} and differs depending on the type of layers contained in the network. 

We will now present the rules used in order to analyze the LocCNN and SampleCNN models.
Specifically, for the input convolutional layer of both networks, we use the $w^2$-rule, chosen since it was proposed for real-valued unbounded inputs~\cite{montavon2017explaining}, as is the case of raw audio signals.
Following the literature~\cite{cho2020layer}, we apply to the 1D convolutional layers the $\gamma$ rule, which favors the effect of positive against negative contributions and is therefore well-suited to the output of ReLU and Sigmoid activations.
We then apply the $\epsilon$ rule~\cite{bach2015pixel} to the fully connected layers.
We disregard the contribution due to the dropout layer and to the activations since these are all monotonically increasing~\cite{perotin2019crnn}. Following~\cite{pahde2023optimizing} we use a custom canonizer to handle the skip connections in SampleCNN and we apply the signal-takes-it-all rule~\cite{arras2017relevant} to handle gating functions in the squeeze-and-excitation boxes.

\begin{figure}[b]
\centering
\begin{minipage}[c]{0.49\columnwidth}
  \centering
\centerline{\includegraphics[width=\columnwidth]{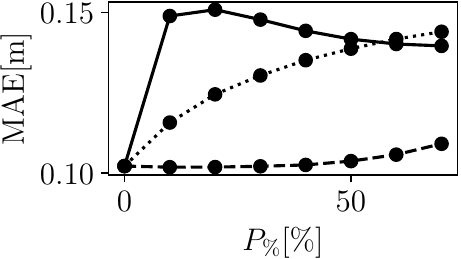}}  \centerline{\footnotesize(a) LocCNN}
\end{minipage}
\hfill
\begin{minipage}[c]{0.49\columnwidth}
  \centering
\centerline{\includegraphics[width=\columnwidth]{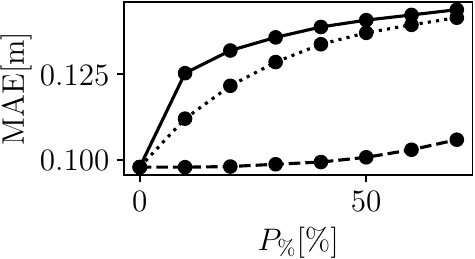}}  \centerline{\footnotesize(b) SampleCNN}
\end{minipage}
\caption{Manipulation of relevant input features results using the random \sampleline{dashed}, amplitude \sampleline{dotted} and LRP \sampleline{} strategies.
}
\label{fig:input_manipulation}
\end{figure}
\begin{figure}[t]
\centering
\begin{minipage}[c]{.3\columnwidth}
  \centering
\centerline{\includegraphics[width=\columnwidth]{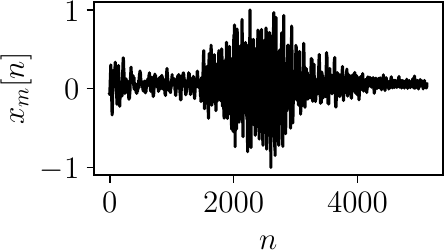}}  \centerline{(a)}
\end{minipage}
\hfill
\begin{minipage}[c]{.3\columnwidth}
  \centering
\centerline{\includegraphics[width=\columnwidth]{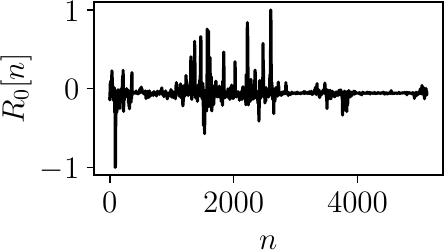}}  \centerline{(b)}
\end{minipage}
\hfill
\begin{minipage}[c]{.3\columnwidth}
  \centering
\centerline{\includegraphics[width=\columnwidth]{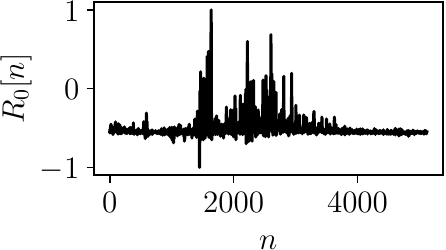}}  \centerline{(c)}
\end{minipage}
\caption{Signal at a microphone placed in $[\SI{0.575}{\metre},\SI{7}{\metre},\SI{1.2}{\metre}]^T$ (a), and corresponding relevance using LocCNN (b) and SampleCNN (c) for a source placed in   $[\SI{1.53}{\metre},\SI{5.71}{\metre},\SI{1.15}{\metre}]^T$.}
\label{fig:example_input}
\end{figure}
\begin{figure*}[t]
\centering
\vfill
\begin{minipage}[c]{0.44\columnwidth}
  \centering
\centerline{\includegraphics[width=\columnwidth]{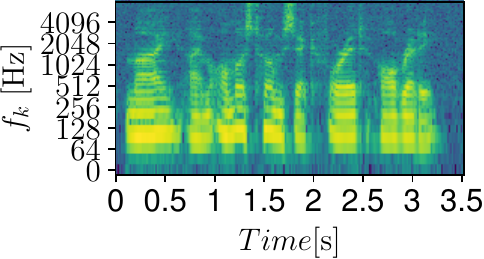}}
\end{minipage}
\hfill
\begin{minipage}[c]{0.44\columnwidth}
  \centering
\centerline{\includegraphics[width=\columnwidth]{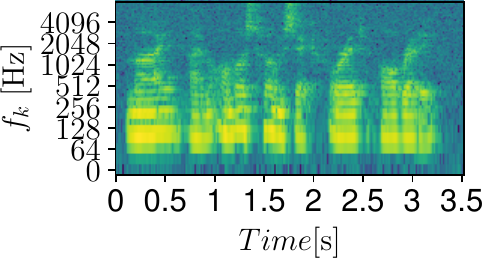}}
\end{minipage}
\hfill
\begin{minipage}[c]{0.44\columnwidth}
  \centering
\centerline{\includegraphics[width=\columnwidth]{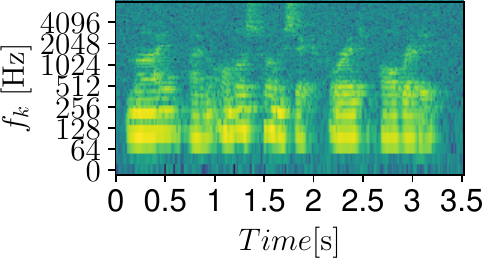}}
\end{minipage}
\hfill
\begin{minipage}[c]{0.44\columnwidth}
  \centering
\centerline{\includegraphics[width=\columnwidth]{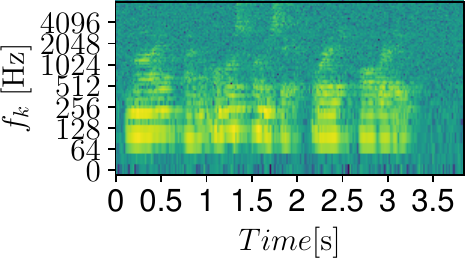}}
\end{minipage}
\vfill
\begin{minipage}[c]{0.44\columnwidth}
  \centering
\centerline{\includegraphics[width=\columnwidth]{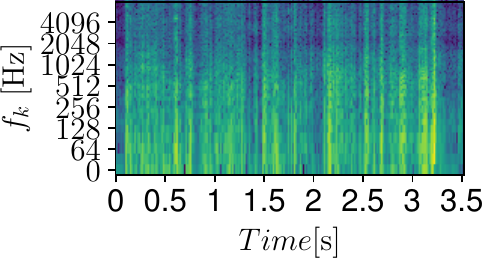}}
\end{minipage}
\hfill
\begin{minipage}[c]{0.44\columnwidth}
  \centering
\centerline{\includegraphics[width=\columnwidth]{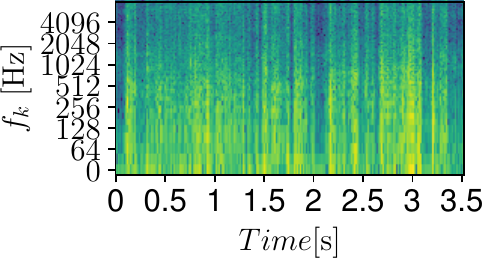}}
\end{minipage}
\hfill
\begin{minipage}[c]{0.44\columnwidth}
  \centering
\centerline{\includegraphics[width=\columnwidth]{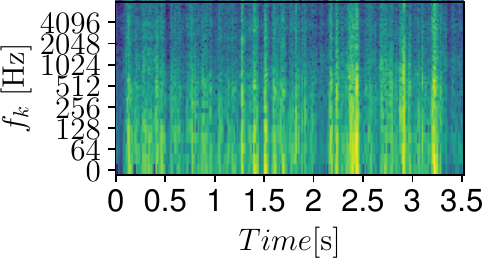}}
\end{minipage}
\hfill
\begin{minipage}[c]{0.44\columnwidth}
  \centering
\centerline{\includegraphics[width=\columnwidth]{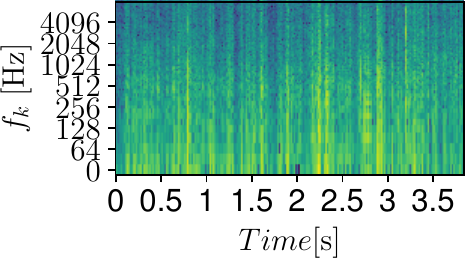}}
\end{minipage}
\vfill
\begin{minipage}[c]{0.44\columnwidth}
  \centering
\centerline{\includegraphics[width=\columnwidth]{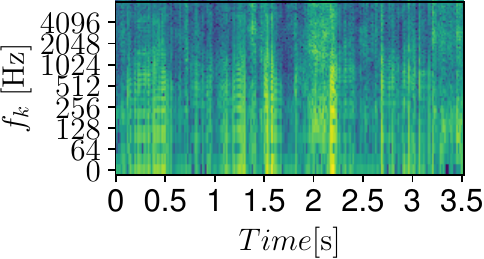}}
  \centerline{\footnotesize(a) $\mathrm{SNR} = 25 \mathrm{dB},T60=0.15~\mathrm{s}$}
\end{minipage}
\hfill
\begin{minipage}[c]{0.44\columnwidth}
  \centering
\centerline{\includegraphics[width=\columnwidth]{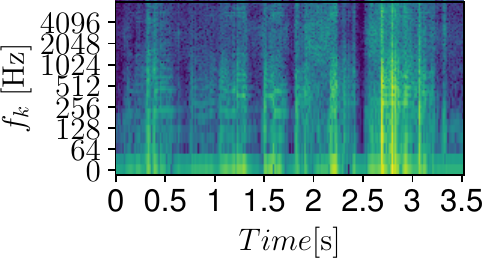}}
  \centerline{\footnotesize(b) $\mathrm{SNR} = 20 \mathrm{dB},T60=0.3~\mathrm{s}$}
\end{minipage}
\hfill
\begin{minipage}[c]{0.44\columnwidth}
  \centering
\centerline{\includegraphics[width=\columnwidth]{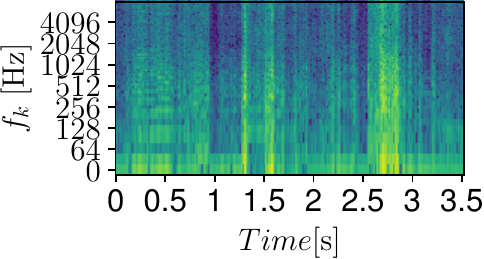}}
  \centerline{\footnotesize(c) $\mathrm{SNR} = 15 \mathrm{dB},T60=0.4~\mathrm{s}$}
\end{minipage}
\hfill
\begin{minipage}[c]{0.44\columnwidth}
  \centering
\centerline{\includegraphics[width=\columnwidth]{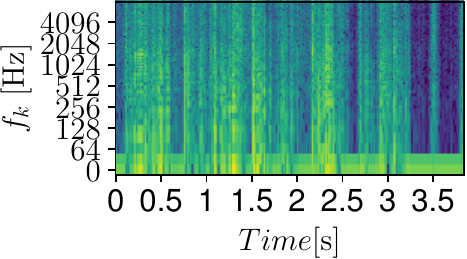}}
  \centerline{\footnotesize(d) $\mathrm{SNR} = 10 \mathrm{dB},T60=0.6~\mathrm{s}$}
\end{minipage}
\caption{STFT of the signal measured at microphone placed in $[\SI{0.57}{\metre}, \SI{7}{\metre}, \SI{1.2}{\metre} ]^T$ from a source placed in $[\SI{1.48}{\metre}, \SI{5.37}{\metre},\SI{1.33}{\metre}]^T$. Top row: microphone signals. Middle row: LocCNN relevance signals. Bottom row: SampleCNN relevance signals. Sub-captions denote corresponding environmental conditions.}
\label{fig:input_visualization_stft}
\end{figure*}
\begin{figure*}[t]
\centering
\begin{minipage}[c]{0.44\columnwidth}
  \centering
\centerline{\includegraphics[width=\columnwidth]{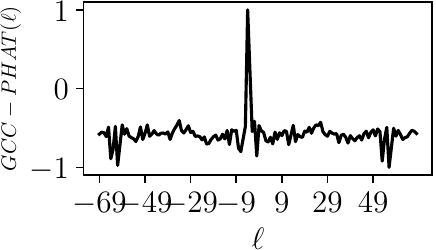}}
\end{minipage}
\hfill
\begin{minipage}[c]{0.44\columnwidth}
  \centering
\centerline{\includegraphics[width=\columnwidth]{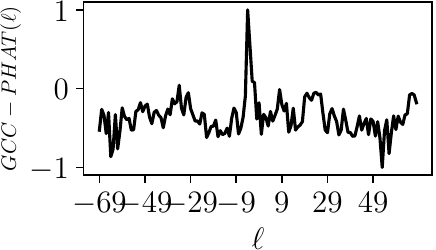}}
\end{minipage}
\hfill
\begin{minipage}[c]{0.44\columnwidth}
  \centering
\centerline{\includegraphics[width=\columnwidth]{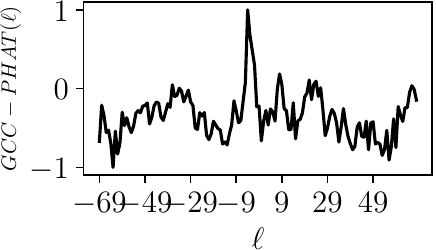}}
\end{minipage}
\hfill
\begin{minipage}[c]{0.44\columnwidth}
  \centering
\centerline{\includegraphics[width=\columnwidth]{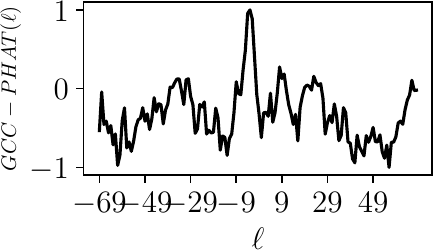}}
\end{minipage}
\vfill
\begin{minipage}[c]{0.44\columnwidth}
  \centering
\centerline{\includegraphics[width=\columnwidth]{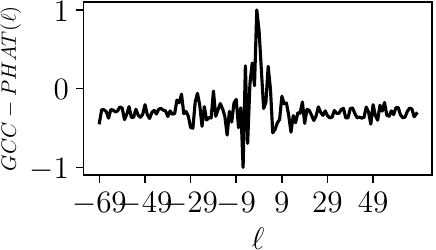}}
\end{minipage}
\hfill
\begin{minipage}[c]{0.44\columnwidth}
  \centering
\centerline{\includegraphics[width=\columnwidth]{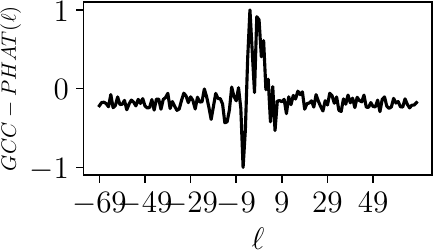}}
\end{minipage}
\hfill
\begin{minipage}[c]{0.44\columnwidth}
  \centering
\centerline{\includegraphics[width=\columnwidth]{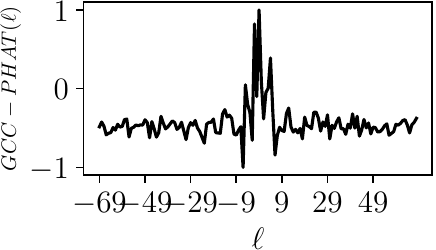}}
\end{minipage}
\hfill
\begin{minipage}[c]{0.44\columnwidth}
  \centering
\centerline{\includegraphics[width=\columnwidth]{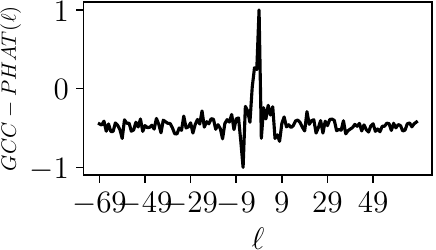}}
\end{minipage}
\vfill
\begin{minipage}[c]{0.44\columnwidth}
  \centering
\centerline{\includegraphics[width=\columnwidth]{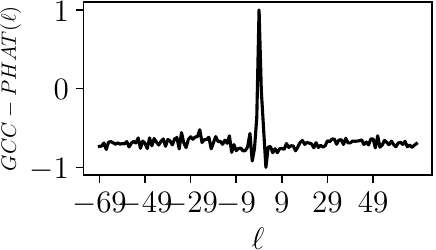}}
  \centerline{\footnotesize(a) $\mathrm{SNR} = 25 \mathrm{dB},T60=0.15~\mathrm{s}$}
\end{minipage}
\hfill
\begin{minipage}[c]{0.44\columnwidth}
  \centering
\centerline{\includegraphics[width=\columnwidth]{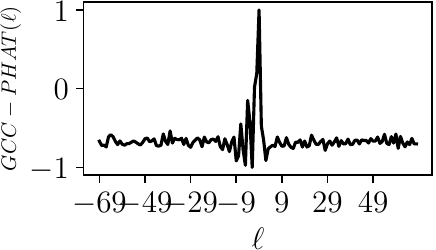}}
  \centerline{\footnotesize(b) $\mathrm{SNR} = 20 \mathrm{dB},T60=0.3~\mathrm{s}$}
\end{minipage}
\hfill
\begin{minipage}[c]{0.44\columnwidth}
  \centering
\centerline{\includegraphics[width=\columnwidth]{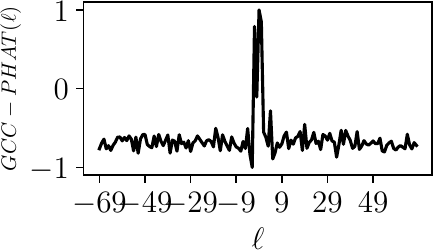}}
  \centerline{\footnotesize(c) $\mathrm{SNR} = 15 \mathrm{dB},T60=0.4~\mathrm{s}$}
\end{minipage}
\hfill
\begin{minipage}[c]{0.44\columnwidth}
  \centering
\centerline{\includegraphics[width=\columnwidth]{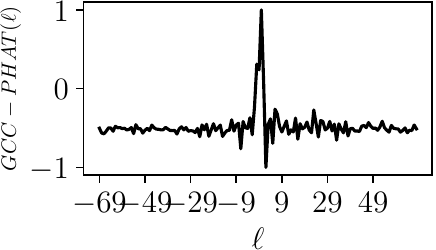}}
  \centerline{\footnotesize(d) $\mathrm{SNR} = 10 \mathrm{dB},T60=0.6~\mathrm{s}$}
\end{minipage}
\caption{GCC-PHATs between two microphones placed in $[\SI{1.47}{\metre}, \SI{7}{\metre},\SI{1.2}{\metre}]^T$ and $[\SI{1.92}{\metre}, \SI{7}{\metre},\SI{1.2}{\metre}]^T$. Source in $[\SI{1.48}{\metre}, \SI{5.43}{\metre},\SI{1.23}{\metre}]^T$. Top row: microphone signals. Middle row: LocCNN relevance signals. Bottom row: SampleCNN relevance signals. 
$\ell$ denotes the time lag in samples. Sub-captions indicate the environmental conditions.}
\label{fig:input_visualization_gcc}
\end{figure*}
\section{Speech Source Localization Models and Environmental Setup}
\label{sec:setup}
In this section we describe how source localization is formulated in the LocCNN and SampleCNN models, provide a brief overview of their architectures, and describe how the application environment was simulated.
\subsection{Signal Model and Problem Formulation}
Let us consider an array of $M$ microphones and a source located in $\mathbf{s}=[s_x,s_y,s_z]^T$ emitting a signal $s[n]$, where $n$ denotes the discrete time index. The signal acquired by the $m$-th microphone can be described as
\begin{equation}
    x_m[n] = h_m[n]*s[n]+e_m[n],
\end{equation}
where $*$ denotes convolution, $h_m$ is the Room Impulse Response (RIR) between $\mathbf{m}$ and $\mathbf{s}$ and $e_m$ an additive noise term.
If we denote $\mathbf{x}_m =[x_m[0],x_m[1],\ldots,x_m[N-1]]^T$ the vector containing the signal acquired by the $m-th$ microphone, where a $N$-sample windowing has been applied and we stack them into the matrix $\mathbf{X}=[\mathbf{x}_1,~ \mathbf{x}_2,~ \ldots,~ \mathbf{x}_M ]\in \mathbb{R}^{N\times M}$, then the source localization task performed by LOC-CNN can be defined as finding the function $f$ such that $\hat{\mathbf{s}} = f(\mathbf{X})$, where $\hat{\mathbf{s}} $ is an estimate of the location of $\mathbf{s}$ in cartesian coordinates.
\subsection{Network Architectures}
The LocCNN architecture consists of $5$ consecutive 1D convolutional blocks, alternated with MaxPooling. The SampleCNN architecture is instead more complex and consists of $5$ residual squeeze-and-excitation boxes. We slightly modified it by adding ReLU and sigmoid activations which showed better performances and were prescribed in the paper where the SampleCNN architecture was first proposed~\cite{lee2018samplecnn}. We also removed the input normalization layer. To compare it with LocCNN we modified its output layer using a fully connected layer with $3$ neurons since the network was originally proposed for a multi-source scenario. Full details available in~\cite{vera18} and~\cite{sundar2020raw}.


\subsection{Setup}
We consider a $3.6~m \times 8.2~m \times 2.4~m$ room where we deploy a uniform linear array (ULA) composed of $M = 16$ microphones, with inter-microphone distance $0.16~m$ at a height of $1.2~\mathrm{m}$. Externally positioned with respect to the array we position $3940$ sources on a $1.75~\mathrm{m}\times 2~\mathrm{m}$ grid along the x-y axes (width-length) and vary their z (height) coordinate by randomly sampling it from a uniform distribution in the interval of $[1~\mathrm{m},1.5~\mathrm{m}]$. We split the dataset using $3152$, $788$, and $288$ samples for train, validation, and test datasets, respectively. The test sources are placed on a $0.5~\mathrm{m}\times 0.5\mathrm{m}$ grid placed in the middle of the whole sources grid.
We consider speech signals extracted from the CMU ARCTIC database \cite{kominek2004cmu}, synchronously sampled at $F_s = 16 \mathrm{kHz}$, using a different recording for each source and simulate the propagation to the microphones using the gpuRIR library~\cite{diaz2021gpurir}.
The environmental conditions of the room are varied considering the following reverberation times ($\mathrm{T60}$) $[0.15~\mathrm{s}, 0.3~\mathrm{s},0.4~\mathrm{s},0.6~\mathrm{s}]$ and Additive White Gaussian Noise (AWGN) with a Signal-to-noise ratio ($\mathrm{SNR}$) of $[10~\mathrm{dB},15~\mathrm{dB}, 20~\mathrm{dB}, 25~\mathrm{dB}]$.
The inputs to both localization techniques were computed by applying a $320~ms$ rectangular window to the microphone signals.
\subsection{Network Training}
We trained both models with a batch size of $100$, following~\cite{vera18} and the Mean Squared Error (MSE) as a loss function. We used the Adam optimizer with a learning rate of $0.001$ for LocCNN and $0.01$ for SampleCNN, which were halved after $100$ consecutive epochs with no validation loss improvement. We set $1000$ as the maximum number of epochs and ended the training when no validation loss improvement was present for $200$ consecutive epochs.

\begin{table*}[t] 
\caption{TDoA estimation, probability of Anomalous Estimates for different microphone spacings.}
\centering
\resizebox{\linewidth}{!}{%
\begin{tabular}{c|c||ccc||ccc||ccc}
\toprule
\multicolumn{2}{c}{} & \multicolumn{3}{c}{$d=\SI{0.15}{\metre}$} & \multicolumn{3}{c}{$d=\SI{0.45}{\metre}$} & \multicolumn{3}{c}{$d=\SI{0.75}{\metre}$}\\
\cmidrule(l){3-5} \cmidrule(l){6-8} \cmidrule(l){9-11}
SNR & T60 & Signal & LocCNN& SampleCNN& Signal & LocCNN& SampleCNN&  Signal & LocCNN& SampleCNN\\ 
\midrule
$25~\mathrm{dB}$& $0.15~\mathrm{s}$ &  \SI{2.64}{\percent}&\SI{0.0}{\percent}&\SI{0.29}{\percent} &\SI{3.0}{\percent}&\SI{3.0}{\percent}&\SI{5.0}{\percent}&\SI{3.25}{\percent}&\SI{21.05}{\percent}&\SI{14.87}{\percent}\\
$20~\mathrm{dB}$ & $0.3~\mathrm{s}$ & \SI{6.39}{\percent}&\SI{0.17}{\percent}&\SI{0.21}{\percent}&\SI{8.0}{\percent}&\SI{4.0}{\percent}&\SI{3.0}{\percent}&\SI{11.41}{\percent}&\SI{23.43}{\percent}&\SI{14.84}{\percent}\\
$15~\mathrm{dB}$& $ 0.4~\mathrm{s}$ & \SI{13.0}{\percent}&\SI{10.0}{\percent}&\SI{3.0}{\percent}&\SI{13.0}{\percent}&\SI{10.0}{\percent}&\SI{3.0}{\percent}&\SI{20.69}{\percent}&\SI{21.46}{\percent}&\SI{17.06}{\percent}\\
$10~\mathrm{dB}$& $0.6~\mathrm{s}$ & \SI{15.98}{\percent}&\SI{1.04}{\percent}&\SI{0.22}{\percent}&\SI{27.0}{\percent}&\SI{7}{\percent}&\SI{3.0}{\percent}&\SI{34.57}{\percent}&\SI{17.96}{\percent}&\SI{23.89}{\percent}\\

\bottomrule
\end{tabular}
}
\label{tab:tde_est}
\end{table*}

\section{Experiments}
\label{sec:experiments}
In this section, we present results that aim to explain the learning process of LocCNN and SampleCNN. Both networks were implemented in PyTorch, while LRP via Zennit~\cite{anders2021software}. 
 Localization error in terms of Mean Absolute Error (MAE) is comprised between $0.088~\mathrm{m}$, in the simplest scenario and $0.11~\mathrm{m}$ for the most challenging one, using both LocCNN and SampleCNN.
 Results are computed over all windowed signals for each test source, for a total of $2911$ examples.
The code, and model weights as well as additional results are publicly available at \url{https://lucacoma.github.io/XAISrcLoc/}.  
\subsection{Manipulation of relevant input features}
Following~\cite{becker2024audiomnist}, we first perform a sanity check aimed at assessing the reliance of the LocCNN and SampleCNN models on features deemed as relevant by LRP, before delving into further analyses.
We apply three different strategies to manipulate a fraction of the input signals by zeroing out chosen samples. We use the \textit{random} strategy as a baseline, where we put samples to zero following a uniform distribution. Then we apply the \textit{amplitude} strategy, where samples are zeroed out according to the absolute amplitude value (i.e. first zeroing samples corresponding to higher absolute amplitude values of the microphone signal), finally, we consider the \textit{LRP} strategy, where samples are zeroed according to the maximal relevance values indicated by LRP. The idea behind this procedure is that, if the network is actually relying on samples deemed relevant by LRP, then the performance should deteriorate more for smaller fractions of manipulation in the case of the LRP strategy, with respect to the others. 

We present in Fig.~\ref{fig:input_manipulation} manipulation results averaged over the $4$ environmental conditions taken into account, by removing samples considering percentages going from $0\%$ up to $70\%$. 
It is clear that using the LRP strategy the MAE increases more steadily with respect to other strategies and performance decrease becomes comparable to the amplitude strategy only when zeroing $60\%$ of samples, this is reasonable, since considering such short windowed signals, at that point, the audio would almost contain only noise. These results show that the samples indicated as relevant by LRP are also relevant to the network and allow us to proceed in further in our study.
\subsection{Input features visualization}
We continue our study by visually inspecting the input provided to the network and the relative relevance signal. We show in Fig.~\ref{fig:example_input}(a) one window of the signal acquired by a microphone and in Fig.~\ref{fig:example_input}(b) Fig.~\ref{fig:example_input}(c) the corresponding relevances obtained via LocCNN and SampleCNN, respectively. As expected, from a simple visual inspection of the 1D signals it is hardly possible to draw any conclusions.

In order to understand if it is possible to derive some visual explanations from the time-frequency representation for the input signal we compute the Short Time Fourier Transform (STFT) of the whole whole speech signal emitted by the source, obtained by concatenating the different windows in which it was divided before inputting it into the network. We using $512$ frequency points, Hann window of length $512$ samples, and hop size of $128$ samples. We show the STFT for a single source and microphone in Fig.~\ref{fig:input_visualization_stft} for all considered environmental conditions computed using the microphone (top row) and LocCNN (middle row) and SampleCNN relevance (bottom row) signals. From the inspection of the STFT we can speculate that the relevance does not preserve the phonemes of the input speech signal\footnote{Listening examples available in the accompanying website.}, but instead simply the pauses between the emission of such phonemes, i.e. corresponding to the temporal contour, suggesting the reasoning that, when performing single source speech localization, the network focuses more on the time at which the signals are emitted than on the speech content. This behavior seems consistent for all noise/reverberation combinations considered and is more evident for the SampleCNN model.

In fact, all these considerations are expected. From a physical understanding of sound propagation, there is no inherent reason why individual samples should be more relevant than others. The information related to the source-sensor geometry can instead be extracted by analyzing the correlations between individual signals acquired by different microphones. To understand if this is the case also for the considered models, we compute the Generalized Cross Correlation with Phase Transform (GCC-PHAT)~\cite{knapp1976generalized} considering two microphones spaced by $\SI{0.45}{\metre}$ as shown in Fig~\ref{fig:input_visualization_gcc} using the concatenated microphone signals (top row), and the concatenated relevance signals corresponding to LocCNN (middle row) and SampleCNN (bottom row) for all considered environmental conditions.
From a simple visual inspection, it seems that the GCC-PHAT obtained via the relevance signals contains fewer spurious peaks, enabling to estimate more easily the TDoA between the sensors and then localizing the sources. This becomes more noticeable as the reverberation and noise conditions worsen. 
But is this actually what the network is learning? In order to try to answer this question we perform a further experiment assessing the performance of the TDoA estimation.
\subsection{TDoA estimation}
We consider three couples of microphones, centered in the middle of the array, with a spacing of $d=0.15~\mathrm{m}$, $d=0.3~\mathrm{m}$, $d=0.75~\mathrm{m}$, respectively. For each couple, we compute the TDoA as the maximum peak of the GCC-PHAT, for all the considered test sources and time windows, for a total of $2911$ examples for each spacing. 
We quantify the TDoA estimation performance in terms of the probability of anomalous estimates $P_{a}$~\cite{comanducci2020time}.
Following~\cite{ianniello1982time} we consider as anomalous estimates the ones exceeding half of the Signal Correlation Time, defined as the width of the main lobe of the autocorrelation function at $\SI{-3}{\decibel}$, for each of the considered speech signals. We compute the GCC-PHAT using both the signals arriving at the microphones and the corresponding relevance. Results are shown in Table~\ref{tab:tde_est}.
The results show that the TDoA estimation performance worsens in all cases with the worsening of the environmental conditions and the increase of the microphone spacing. However, the error increases using the measured signals is more steep with respect to the one obtained using the signals obtained via LRP. The relevance signals indicate which part of the input signals are deemed important by the networks to estimate the source location. This means that both models learn from the microphone signal information that enables to better estimate of the TDoA, suggesting, that the networks leverage on statistical correlation between the microphone signals to estimate the source position. 

\section{Conclusion}
\label{sec:conclusion}
In this paper, we studied the learning process of two end-to-end deep learning models, differing in complexity, for speech source localization, mapping raw multichannel audio signals to cartesian positions. We applied the Layer-wise Relevance Propagation technique to understand which parts of the input are more important in determining the predicted output. The performed analyses suggest that the networks learn to denoise and de-reverberate the signals in order to better correlate them and consequently estimate the TDoA. We believe that the obtained results should contribute to motivating the adoption of XAI techniques as a standard practice in deep learning applications for acoustic signal processing problems.
\ifCLASSOPTIONcaptionsoff
  \newpage
\fi
\balance
\bibliographystyle{IEEEtran}
\bibliography{biblio}

\end{document}